\documentstyle[epsf]{elsart}

\begin{document}
\begin{frontmatter}
\title{Learning a spin glass:\\ determining Hamiltonians from metastable states.}
\author{S. M. Kuva\thanksref{cnpq}},
\author{O. Kinouchi\thanksref{fapesp}} and 
\author{N. Caticha\thanksref{recope}\thanksref{mail}}
\address{ Departamento de F\'{\i}sica Geral, Instituto de F\'{\i}sica \\
Universidade de S\~ao Paulo \\
Caixa Postal 66318,  
CEP 05315-970, S\~ao Paulo, SP, Brazil.}
\thanks[cnpq]{Supported by CNPq and FAPESP.}
\thanks[fapesp]{Supported by FAPESP.}
\thanks[recope]{Partially supported by CNPq and FINEP/RECOPE.}
\thanks[mail]{Corresponding author: nestor@gibbs.if.usp.br}

\begin{abstract}
We study the problem of determining the Hamiltonian of a fully connected
Ising Spin Glass of $N$ units from a set of measurements, whose sizes needs
to be ${\cal O}(N^2)$ bits. The student-teacher scenario, used to study learning
in feed-forward neural networks, is here extended to 
spin systems with arbitrary couplings.
The set of measurements consists of data about the local
minima of the rugged energy landscape.
We compare simulations and analytical approximations for the resulting 
learning curves obtained by using different algorithms.

\end{abstract}

 
\end{frontmatter}

\section{Introduction}

The study of the dynamics or statistical properties of a system
usually consists in making predictions of its behavior based on
assumed microscopic 
laws such as, for example, using knowledge about its Hamiltonian. 
However, ill posed and inverse problems can be found in a vast 
array of areas. Typically, in these cases, the problem is not to
find the behavior, but rather to obtain the microscopic laws 
that gave rise to it. Among the many interesting questions that
can be asked, we point out those about the structure of the law,
its uniqueness  and how well it can be determined based on 
partial information.

Inverse problems of different degrees of difficulty that have
been subject of recent intense research activity include rule
extraction and learning in artificial systems; pattern recognition,
clustering and categorization problems; 
to find out the
sequence of amino acids that leads to a predetermined chemical
activity;  obtaining the parameters of a dynamical system from
the time series it generates; obtaining renormalized Hamiltonians
from Monte Carlo Renormalization Group data etc.

In dealing with these problems, techniques from statistics, 
combinatorial optimization, statistical mechanics, dynamical
theory and other areas have been found useful to different degrees.
In this paper, we deal with the problem of determining the 
Hamiltonian of a spin glass from data about its metastable states
(MS), that is, {\em learning} a spin glass. Related issues has
been recently addressed by Kanter and Gotesdyner \cite{KG}. 
In particular, they ``show(ed) that static properties determine
the dynamics for a large class of systems'' and asked whether
``Classical spin systems with the same MS have the same 
Hamiltonians''. The affirmative answer, for a large class of systems,
immediately calls for the following question --- how hard is it
to determine the Hamiltonian from partial information about the MS?
In trying to answer this, we use ideas from learning in neural
networks \cite{WRB}. 

\section{On-line learning in a Spin Glass System}

A set of MS is used as a learning set in the
{\em student-teacher} scenario. The teacher being the original
classical fully connected spin system of Hamiltonian ${\cal H}_t = \frac{1}{N}
\sum B_{ij}S_i S_j$. The student, another system of a similar
structure ${\cal H}_s = \frac{1}{N}\sum J_{ij}S_i S_j$, is our approximation
for the teacher, being constructed
from the MS data.

Here we specialize to the case of an Ising Spin Glass teacher whose couplings 
are drawn independently from the distribution 
$P(B_{ij})= \frac{1}{2}\delta(B_{ij}-1) + \frac{1}{2}\delta(B_{ij}+1)$.
The self-couplings $B_{ii}$ are set to zero.
The training set is generated by letting a randomly chosen
initial teacher configuration ${\boldmath S}_0$  relax to the
nearest local minimum  ${\boldmath S}^*\equiv 
{\boldmath S}(t\rightarrow \infty) $
according to a zero temperature aligning-field dynamics,
\begin{equation}
S_i(t+1) = \mbox{sgn} \left(\sum_{j \neq i}B_{ij}S_j(t)\right) 
\stackrel{t\rightarrow \infty }{\longrightarrow}S^*_i \:.
\end{equation}

Both teacher and student spin systems are equivalent to $N$ fully 
connected perceptrons. The $i$-th student perceptron
learns from a set of $\nu=1,\ldots,p=\alpha N$
examples, ${\cal L}_i=\{\mbox{\boldmath S}^{i\nu},\sigma_i^{\nu}\}$.
The input vector
\begin{equation} 
\mbox{\boldmath S}^i = \{S^*_1,\ldots,S^*_{i-1},0,
S^*_{i+1},\ldots,S^*_N\}
\end{equation} 
for site $i$ is obtained from the metastable configuration by
setting to zero the $i$-th component; that component is the desired output
$\sigma_i\equiv S^*_i$.
The task of the learning process is to build a
student with the same energy landscape
of the teacher system or, equivalently, to estimate the 
Hamiltonian parameters $\{B_{ij}\}$. We show that it is possible
using only information contained in a small set of MS (${\cal O}(N)$) in
comparison to the exponential number of spin-glass local minima
($\exp(0.19 N)$ \cite{KG,FH}).

In the online learning \cite{KC92,CKC} strategy,
each example is presented only once inducing a change in the synaptic weights
in the following way,
\begin{equation}
J_{ij}(\nu+1) = J_{ij}(\nu)+\frac{1}{N}F_i(\nu)\sigma_i^{\nu}S_j^{*\nu}
\:\:\:\:\mbox{(for $i\neq j$)}\:. 
\label{update}
\end{equation}
The function $F_i(\nu)$ modulates the Hebb term 
$\sigma_i^{\nu} S_j^{*\nu}$ and characterizes different learning
algorithms. The self-couplings are always set to $J_{ii}=0$.

The macroscopic description of the learning process involves the
quantities $Q_i= \sum_j^N J_{ij}J_{ij}/N$
(the squared norm of the $i$-th student perceptron), $M_i= 
\sum_j^N B_{ij}B_{ij}/N$ (the corresponding teacher squared norm).
We also define the normalized teacher-student overlap 
\begin{equation}
\rho \equiv\frac{1}{N}\sum_i \rho_i = \frac{1}{N}\sum_i^N \sum_j^N
\frac{B_{ij}J_{ij}}{\sqrt{M_iQ_i}}\:,
\end{equation}
which will be our performance measure as a function of the number of
examples $\alpha N$. The normalized teacher and student local fields
at site $i$ are $b_i\equiv \sum_j^N B_{ij} S_j/\sqrt{B_iN}$ 
and $h_i \equiv \sum_j^N J_{ij} S_j/\sqrt{Q_iN}$. 

In the simpler case of feed-forward networks, 
the update rule Eq.~(\ref{update}) leads,
in the thermodynamic limit, to a set of coupled differential 
equations describing the
order parameters learning dynamics \cite{CKC}. 
In order to proceed and obtain similar equations
we need to make several approximations which will eventually be checked
by simulations. First, we ignore the correlation among the 
minima and thereby do not take into
account the effects that different choices of the 
particular sequence of MS will have.
Second, we assume self-averaging of the order parameters
and site symmetric evolution ($\rho_i(\alpha)=\rho(\alpha)$). 
Finally, we assume that the relevant features of $P(\mbox{\boldmath S})$
can be incorporated into an uncorrelated 
teacher local field distribution $P(\{b_i\})=\prod_i P(b_i)$.
This obviously is not the case in the spin-glass problem. 
Even if the true
teacher local fields distribution $P(b)$ was known, the distribution of
local minima $P(\mbox{\boldmath S}^*)$ 
has special directions related in a complicated way to the
vectors ${\boldmath B}_i=\{B_{ij}\}_{j=1,\ldots,N}$. 
It will be interesting, however, to compare the performance
in the spin glass problem with the theoretical results for a simple approximation
for the distribution of local fields suggested 
by Palmer and Pond \cite{PP}: 
$P(b)=(2s^2)^{-1} |b|\exp\left(-\frac{1}{2}b^2/s^2\right)$.
Although this has been proposed for the fields of global minima states, it is also a good approximation for local minima fields \cite{KG}. The parameter $s$
is adjustable, and in our simulations we obtained $s \approx 1.05$.

Within these approximations, instead of $2N$ learning 
equations (two for each site), we need only two:
\begin{eqnarray}
\frac{d\rho}{d\alpha}&=&\left< \sigma\left(b-\rho h\right)
\frac{F}{\sqrt Q}
-\frac{\rho}{2}\frac{F^2}{Q} \right>\:, \nonumber \\
\frac{dQ}{d\alpha}&=&2Q\left< \sigma h \frac{F}{\sqrt Q} + 
\frac{F^2}{2Q} \right>\:, \label{dyn}
\end{eqnarray}
where $\left< ... \right> = \int db \:dh\: (...) P(b,h)$. The joint density
$P(b,h)$ can be written as $P(h|b) P(b)$, with 
$P(h|b)=\left[2\pi(1-\rho^2)\right]^{-1/2} \exp\left[-
\frac{1}{2}(h-\rho b)^2/(1-\rho^2) \right]$ and $P(b)$ being the particular
teacher fields distribution to be studied.

The asymptotic $(\alpha \rightarrow \infty)$ 
behavior of the `order' parameter $\rho$ is a quality measure for
comparing the algorithms. In this limit we write
\begin{equation}
 1-\rho(\alpha) \approx C \alpha^{-2x}\:,
\end{equation}
where $x$ is the usual {\em learning exponent\/} considered in the literature.

\section{Results}

The internal fields in the spin glass examples are correlated in some
unknown way. It is interesting to observe the effect of these correlations
by comparing with the case where the teacher local fields 
are assumed to be independently distributed ($P(\{b_i\}=\prod_i P(b_i)$) 
according to
\begin{equation}
P_{(r,s)}(b) = \frac{|b|^r}{{\cal Z}(r,s)} \exp \left(- \frac{b^2}{2s^2}\right) \:,
\label{ppond}
\end{equation}
parametrized by $r$ and $s$; ${\cal Z}(r,s)$ ensures normalization.
The Palmer--Pond distribution is achieved by setting $r=1$.

The learning equations (\ref{dyn}) are {\em exact\/} in 
the thermodynamic limit if the learning sets ${\cal L}$ were generated
by choosing random independent identically distributed
examples whose teacher fields obey the above 
Palmer--Pond-like distribution (this case will be denoted IID PP).
Standard analytical calculations \cite{CKC} in the $ \rho \rightarrow 1$
limit, using the distribution (\ref{ppond}) 
lead us to the following results:

{\bf Simple Hebb rule}, $F=1$: we obtained $x=1/2$, independent of 
$r$; the prefactors as functions of $r$ and $s$ are given by
\begin{eqnarray}
C(0,s) &=& \frac{\pi}{4s^2} \:,\:\:\:\:C(1,s) =  \frac{1}{\pi s^2}\:, \nonumber \\
C(r,s) &=& \left[\frac{\Gamma \left(\frac{r+1}{2}\right)}{
\Gamma \left(\frac{r+2}{2}\right)} \right]^2 \frac{1}{4 s^2} \:;  
\end{eqnarray}

{\bf Rosemblatt Perceptron algorithm}, $F=\Theta(-\sigma h)$: 
the learning exponent is $x=1/(3+r)$, with the following prefactors:
\begin{eqnarray}
C(0,s) &=& \left(\frac{s}{3\sqrt{2}}\right)^{2/3} \:,\:\:\:\:
C(1,s) = \frac{s}{8}\sqrt{\frac{3 \pi}{2}} \:, \nonumber\\
C(r>0,s)& = & \frac{1}{2} 
\left[\frac{{\cal Z}(r,s) I(r,0)}{4(r+3)I(r,1)I(r+1,0)}
\right]^{2/(r+3)}\:;
\end{eqnarray}
with ${\cal Z}(r>0,s) = 2 s^{r+1}\sqrt{2\pi} I(r-1,0)$ and
$I(r,n)\equiv  \int_0^\infty dz \:z^r\int_z^\infty Du\: u^n $;
$Du\equiv du \exp(-u^2/2) /\sqrt{2\pi}$ and
$\Gamma(x)$ is the Gamma function.
All the $I(r,n)$ integrals can be found by using 
\begin{eqnarray}
I(r,0) & = & \frac{2^{r/2}}{(r+1)\sqrt{2\pi}}
 \:\Gamma \left(\frac{r+2}{2}\right)\:, \nonumber\\
I(r+1,1) & = & (r+1) I(r,0) \:,\\
I(r,n+2)    &=& (r+n+1)I(r+n,0)+(n+1)I(r,n)\:, \nonumber 
\end{eqnarray} 
starting from $I(0,0)=1/\sqrt{2\pi}$ and $I(0,1)=1/2$.  

It is worth to note that, due to the
behavior $x=1/(3+r)$, the Perceptron algorithm shows only partial learning
in the limit $r \rightarrow \infty $ (as indicated by $x \rightarrow 0$),
that is, when $P(b)$ goes to zero exponentially as $b \rightarrow 0$,
(say, as $ \exp(-1/b^2)$).
This condition relates to, but is weaker than, the case of distributions with
a gap around $b=0$ discussed by Reimann and Van den Broeck \cite{RVB}.
Another interesting point is that, since in principle $r$ may be any real non-negative
number, the learning exponent $x$ can assume
real values which are not simple fractions. The previous ubiquity of
these fractions found in the literature reflects simply the particular 
choices for the small $b$ behavior 
of the distributions $P(\mbox{\boldmath $S$})$ studied so far.

\begin{figure}[htb]
\epsfxsize = 0.8\textwidth
\begin{center}
\leavevmode
\epsfbox{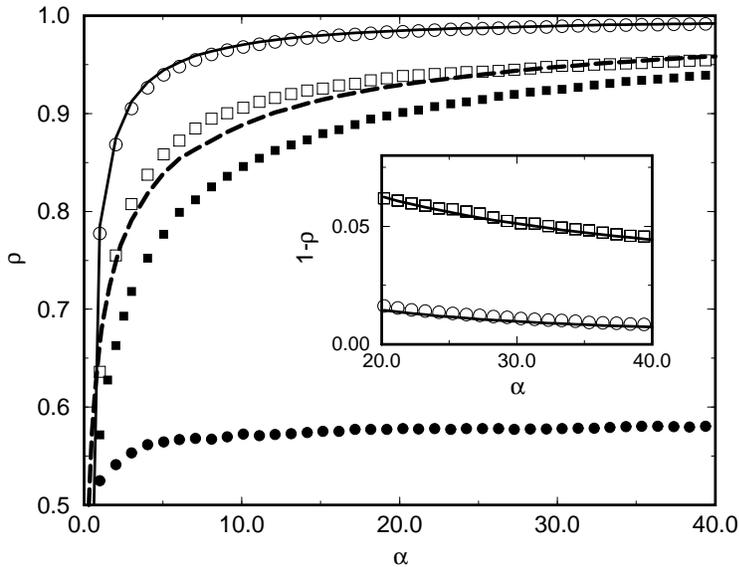}
\end{center}
\caption{Evolution of $\rho$ as a function of $\alpha$. Lower curves are for
a Spin glass teacher ($N=399$): 
Hebb (balls), Perceptron (squares) and $F_{opt}(1,0)$
(dashed). Upper curves
are for the IID PP case ($N=99$): 
Hebb (circles) and Perceptron (squares) and $F_{opt}(0,1)$
(solid). This last curve is slightly above the Hebb curve.
Insert: asymptotical behavior of $1-\rho$ for Perceptron (squares)
and Hebb (circles) for the IID PP case. Solid lines are theoretical curves.}
\label{fig1}
\end{figure}

We compare these analytical results with simulations for examples
generated by the IID PP case with $r=1$ and $s=1.05$, 
and also with simulation results for the spin-glass teacher, see Fig.\ref{fig1}. 
Since the spin glass local minima distribution has structure and special directions
(related in an unknown way with the matrix $\{B_{ij}\}$), we
expect that the results are only approximate.

Indeed, the simulations for examples
distributed exactly according to (\ref{ppond}) are in excellent
agreement with the theoretical predictions,
see Table 1 and Fig.~\ref{fig1} (insert). 
But in the simulations with the spin glass teacher, 
simple Hebb learning stops at
$ \rho(\alpha\rightarrow \infty) \approx 0.57$ ({\em partial learning\/}). 
Analogously to the results of
Riegler {\em et al.\/} \cite{RBSM}, this can be attributed to the presence of
special directions in $P(\mbox{\boldmath $S^*$})$ not aligned with 
${\boldmath B}_i$. 

This lack of robustness is a feature only of the simple Hebb rule.
The Perceptron learning algorithm is robust to 
the correlations in the examples provided by the spin glass teacher: 
perfect learning is achieved in the $\alpha \rightarrow \infty$ limit.
The learning exponent $x$, however, 
seems to be $1/3$ instead of the theoretically expected value $1/4$. This
is a finite size effect which cannot be eliminated by using larger
systems due to the following non-uniform convergence phenomena.
In the Palmer--Pond distribution, 
$P(b)\rightarrow 0$ as $|b|\rightarrow 0$, but in the SG 
simulations, $P(b)$ assumes a finite value $P(0)=a$ of 
order ${\cal O}(1/\sqrt{N})$ at this point due to
finite size effects\footnote{This finite
size effect should not to be confused with the finite $P(0)$ found
in the replica symmetric calculation of Roberts~\cite{Roberts}, 
which is presumably
wrong due to broken replica symmetry.} \cite{PP,BP}. 
A better description of the local field
distribution is obtained by the form
\begin{equation}
P(b) = \frac{a(N)+|b|}{{\cal Z}} \exp \left(- \frac{b^2}{2s^2}\right) \:.
\end{equation}
Repeating the calculation with the above distribution
we found that any finite parameter $a$ changes
the large $\alpha$ behavior, 
leading to $x=1/3$, the same value found for the $r=0$ distribution.
Thus, learning the spin-glass Hamiltonian is
easier for finite $N$.

\begin{table}[htb] 
\vspace{0.5 cm}
\begin{center}
\caption{Learning exponent $x$. The first two columns were obtained by
numerical simulations for increasing $N$ and making an extrapolation to $N\rightarrow
\infty$; the last two, by analytical calculations
using the distribution Eq.~(\ref{ppond}).} 
\vspace{0.5 cm}  
\begin{tabular}{||c|c|c|c|c||} 
\hline
\centering

                  &Spin Glass   & IID PP  &r=0& r=1 \\ \hline\hline
HEBB              &0.001 $\pm$ 0.005    &0.52 $\pm$ 0.01   &1/2 & 1/2 \\ \hline
PERCEPTRON        &0.314 $\pm$ 0.01     &0.245 $\pm$ 0.01    &1/3 & 1/4 \\ \hline

\end{tabular}
\end{center}
\end{table}

\section{Robustness of the `Optimal algorithm'}

The optimal performance for on-line learning in the class of distributions
examined here is given by the prescription \cite{KC92,CKC}
\begin{eqnarray}
F_{opt}^{(r,s)}(\sigma,h) & = & \sqrt{Q} \left( \rho^{-1} 
\sigma\langle b \rangle_{b|h,\sigma} - \sigma h\right)\:, \label{opt} \\
 & = & - \sqrt{Q}\frac{1-\rho^2}{\rho^2} \sigma \frac{d}{dh} 
\ln \frac{P_{(r,s)}(\sigma,h)}{P_0(h)}\:,\nonumber
\end{eqnarray}
where $\langle(...)\rangle_{b|h,\sigma} \equiv \int db\:(\ldots)\:
P_{(r,s)}(b|h,\sigma)$, 
$\sigma=\mbox{sgn}(b)$ and
$P_0(h)= \e^{-h^2/2}/\sqrt{2\pi}$. The
distribution $P_{(r,s)}(\sigma,h)$ is obtained from the 
joint distribution $P_{(r,s)}(b,h)=
P(h|b) P_{(r,s)}(b)$ by integration over $|b|$, and the distribution
$P_{(r,s)}(b|h,\sigma)$ is equal to $P_{(r,s)}(b,h)/P_{(r,s)}(\sigma,h)$.

Since optimal algorithms are obtained always for specific 
distributions, they could suffer from lack of robustness.
This possibility seems not to be a serious problem in the
absence of specific knowledge of, e.g., $r$ and $s$.
We have done simulations with 
\begin{equation}
F_{opt}^{(0,1)} =\sqrt{Q} \frac{\lambda}{\sqrt{2\pi}} 
\frac{\e^{-h^2/2\lambda^2}}{H(-\sigma h/\lambda)} \:,
\end{equation}
where $\lambda=\sqrt{1-\rho^2}/\rho$ and 
$H(x)=\int_x^\infty Du$,
which is the optimal algorithm
for Gaussian teacher local fields $(r=0)$ with unit variance
\cite{KC92}. The examples, however, are generated with
the Palmer--Pond distribution with parameters $r=1, s=1.05$ and
with the spin glass teacher. 
Although not optimal,
the performance of the $F_{opt}^{(0,1)}$ is better than standard algorithms,
see Fig.~\ref{fig1}.

Thus, although derived for specific distributions of examples, optimal
algorithms can be used successfully for other distributions.
The robustness of the optimal algorithm arises as a very 
welcome property, since in real world problems the
examples may be nontrivially distributed in an unknown manner.

\section{Conclusion}

We studied the learning process in neural networks in a
scenario at the midway between the simple
distributions of examples studied so far in the literature \cite{RBSM,RVB}
and real world problems.
The true distribution $P({\boldmath S})$ of `examples' (local minima)
generated by the spin-glass is {\em unknown\/}, but the teacher
system is yet {\em realizable\/} by the student network. 
We have compared the performance of standard algorithms
in this spin-glass problem
with theoretical and simulation results for examples 
with a Palmer--Pond distribution $P_{(r,s)}(b)$ for
the local fields.

Various extensions on this scenario can be devised: we may study 
teachers with more structured distribution of local
minima such as Hopfield networks. We expect that these Hamiltonians
are harder to learn since they have less local minima. For example,
a ferromagnetic Hamiltonian with only two global minima is unlearnable
because many ${\boldmath J}$ vectors are compatible with these two
`examples' \cite{KG}. Another interesting extension
is to learn from a teacher which generates
examples at a non zero temperature. This corresponds to learning from
`noisy examples' with a noise level depending in a non trivial way on
the temperature $T$. Since it has been demonstrated that it is possible
to learn perfectly from noisy examples \cite{CKC,BRS}, we expect that
this task is also learnable.

Finally, we think that our work opens an unexplored learning scenario 
where the distribution of examples is generated dynamically
by the teacher system but the teacher architecture is yet {\em realizable\/}
by the student. Another example could be the learning from examples
generated from the
attractor time-series of a recurrent perceptron \cite{Kanter,KT}.
It is worthwhile to study these {\em realizable\/} cases
since they define 
{\em upper bounds\/} for the performance achievable by neural
networks. The approach of determining theoretical upper bounds for 
the efficiency of simple (thermal or computational) machines follows a
long tradition in Thermodynamics and Statistical Physics.

\begin{ack}
We thank J. F. Fontanari for discussions about the SG fields
distribution.
\end{ack}


\begin{thebibliography}{99}

\bibitem{KG} I. Kanter and R. Gotesdyner,
{\em Phys. Rev. Lett.\/} {\bf 72} (1994) 2678.

\bibitem{WRB} T. L. H. Watkin, A. Rau and M. Biehl,
{\em Rev. Mod. Phys.\/}, {\bf 65} (1993) 499.

\bibitem{FH} K. H. Fischer and J. A. Hertz,
{\em Spin Glasses\/}, Cambridge University Press, Cambridge, UK, 1991.

\bibitem{KC92} O. Kinouchi and N. Caticha,
{\em J. Phys. A: Math. Gen.\/} {\bf 25} (1992) 6243.

\bibitem{CKC} M. Copelli, O. Kinouchi and N. Caticha,
{\em Phys. Rev. E\/} {\bf 53} (1996) 6341.

\bibitem{PP} R. G. Palmer and C. M. Pond,
{\em J. Phys. F: Metal Phys.\/} {\bf 9} (1979) 1451.

\bibitem{RBSM} P. Riegler, M. Biehl, S. Solla and C. Marangi, in:
{\em Neural Nets WIRN Vietrie-95,
Proceedings of the 7th Italian workshop on neural nets\/},
M. Marinaro and R. Tagliaferri, eds., 
World Scientific, Singapore, 1996.

\bibitem{RVB} P. Reimann and C. Van den Broeck,
{\em Phys. Rev. E\/} {\bf 53} (1996) 3989.

\bibitem{BP} F. T. Bantilan Jr. and R. G. Palmer, 
{\em J. Phys. F: Metal Phys.\/} {\bf 11} (1981) 261.

\bibitem{Roberts} S. A. Roberts, 
{\em J. Phys. C: Solid State Phys.\/} {\bf 14},
3015 (1981). 

\bibitem{BRS} M. Biehl, P. Riegler and M. Stechert,
{\em Phys. Rev. E\/} {\bf 52} (1995) R4624.

\bibitem{Kanter} I. Kanter, A. Kessler,
A. Priel and E. Eisenstein,
{\em Phys. Rev. Lett.\/} {\bf 75} (1995) 2614.

\bibitem{KT} O. Kinouchi and M. H. R. Tragtenberg, 
{\em Int. J. Bifurcation and Chaos\/} {\bf 6} 12A (1996) 2343.


\end{thebibliography}
\end{document}